# VMR-Phase Ⅰ VKW 长短视频可自定义关键词唤醒比赛系统综述

作者姓名(袁有根[1]　吕志强[1]　黄申[1]　胡鹏飞[1])

作者单位全称，省市及邮编 (1.腾讯科技（北京）有限公司，北京，100193)

**摘　要**：关键词唤醒技术一直语音处理的研究热点，但是很多相关工作都是在不同的数据集上做的。我们此次组织了一个汉语长短视频直播语音关键词竞赛(Video Keyword Wakeup Challenge, VKW)，旨在公开数据集下检验各个参赛队伍构建关键词唤醒系统的能力。所有提交的系统不仅需要支持设定多个不同关键词，而且需要支持任意自定义关键词的唤醒。本文主要描述了 VKW 竞赛的基本情况，以及部分参赛队伍的实验结果。

**关键词**：VKW；关键词唤醒；公开测试集

**中图分类号**：　　　　　　　　　　　　　　**文献标识码**：

## VRM-Phase I VKW system description of long-short video customizable keyword wakeup challenge

作者姓名拼音 (Yuan Yougen[1]　Lv Zhiqiang[1]　Huang Shen[1]　Hu Pengfei[1])

作者单位英文名称，省市及邮编，国名(1. Tencent Technology (Beijing) Co., Ltd, Beijing, China, 100193)

**Abstract:** Keyword wakeup technology has always been a research hotspot in speech processing, but many related works were done on different datasets. We organized a Chinese long-short video keyword wakeup challenge (Video Keyword Wakeup Challenge, VKW) for testing the ability of each participating team to build a keyword wakeup system under the public dataset. All submitted systems not only need to support the setting of multiple different keywords, but also need to support the wakeup of any costumed keyword. This paper mainly describes the basic situation of the VKW challenge and the experimental results of some participating teams.

## 1　引言

关键词唤醒技术是指从连续语音流中检出所有目标关键词及其出现位置[1]。与自动语音识别技术相比，它不关注非关键词的语音内容，具有计算量少和推理速度快等优势。因此关键词唤醒技术吸引了大量的研究人员去改进算法[2,3,4]，也被应用到越来越多的实际场景中。

为了公平对比不同的技术算法，很多研究人员都会通常会选择公开测试集或者竞赛进行验证。据我所知，NIST 国际评测是比较有影响力的竞赛，它的 OPENKWS 系列竞赛[5]已经连续举办四届，旨在解决电话 PSTN 信道和低资源小语种条件下的关键词定位和识别能力。但是，在互联网音视频中，和关键词唤醒相关的公开数据集和竞赛依然是一片空白。

---

因此，我们组织了一个汉语长短视频直播语音关键词竞赛(Video Keyword Wakeup Challenge, VKW)，旨在公开数据集下检验各个参赛队伍构建关键词唤醒系统的能力。所有提交的系统不仅需要支持设定多个不同关键词，而且需要支持任意自定义关键词的唤醒。VKW竞赛首先将关键词唤醒任务搬到了互联网音视频中，并且选取了当前工业界最为关注的三类媒体场景--长视频、短视频、直播；然后该竞赛分了受限和非受限两个赛道以供各个参赛队伍选择，并且提供了公开数据集；最后，该竞赛还提供了统一的评测工具来公平衡量不同的关键词唤醒系统。

本文主要描述了VKW竞赛的基本情况，并且展示了部分参赛队伍的实验结果。整体而言，VKW竞赛一共吸引了61个团队报名参加，每个团队有45天去完成关键词唤醒的系统搭建，调试和结果提交工作，最终收到了23套提交系统。很多参赛团队提交的关键词唤醒系统都体现出了非常高的技术水准，部分团队的关键词唤醒系统甚至远超基线。

## 2 竞赛内容

为了更公平的比较各个团队的关键词唤醒系统，此次VKW竞赛将被分成受限赛道和非受限赛道。其中受限赛道的目的是用指定的公开数据集和开源方法去提升关键词唤醒的技术效果，非受限赛道的目的是尽一切可能去提升关键词唤醒技术的最终效果。

### 2.1 受限赛道

在受限赛道下，参赛团队指允许使用指定的内部数据和外部数据。内部数据是指有标注的语音训练数据，只包括官方提供的1505小时普通话朗读数据，长视频、短视频、直播各5小时数据。外部数据是指无标注的开源语音训练数据，也可以对开源语音训练数据进行数据扩充，预训练和语音合成等操作。

另外，参赛队伍也可以直接使用开源的预训练模型、语言模型、语音合成接口和网络爬取的文本等。在构造发音词典过程中，参赛队伍可以直接采用官方提供的标注发音词典，也可以采用公开渠道可获取的词典，包括通过商业渠道或者开源项目提供的发音词典。

### 2.2 非受限赛道

在非受限赛道，参赛队伍可使用可公开获取的标注数据、任意无标注数据进一步提升系统性能，但需要在最终提交系统说明里提供数据来源。

## 3 数据设置

本次VKW竞赛提供1505小时普通话朗读数据作为训练集；长视频、短视频、直播数据各5小时作为微调集；长视频、短视频、直播数据各5小时作为开发集；长视频、短视频、直播数据各20小时为作为测试集。具体情况如下表1所示，其中训练集和微调集可以用来训练模型，开发集仅可用于测试模型效果，不允许用来进行模型训练。测试集用于提交模型结果和统一对比。

表1 VKW公开数据集

| 数据集 | 训练集（小时） | 微调集（小时） | 开发集（小时） | 测试集（小时） |
| --- | --- | --- | --- | --- |
| 朗读 | 1505 | N/A | N/A | N/A |
| 长视频 | N/A | 5 | 5 | 20 |
| 短视频 | N/A | 5 | 5 | 20 |
| 直播 | N/A | 5 | 5 | 20 |

## 4 评价指标

本次 VKW 竞赛的评价指标有两个：F1 和 ATWV(Actual Term-weighted Value)[6]。其中 F1 指标是对全部关键词的整体精准度（Precision）和召回率（Recall）的一个调和平均，它反映了关键词唤醒系统对当前词表的综合性能。F1 指标的计算公式为：

$$\text{Precision} = \frac{N_{Corr}}{N_{Corr} + N_{FA}}$$

$$\text{Recall} = \frac{N_{Corr}}{N_{True}}$$

$$F1 = \frac{2 * Precision * Recall}{(Precision + Recall)}$$

其中，$N_{Corr}$ 代表正确命中关键词的个数，$N_{FA}$ 代表误命中的关键词个数，$N_{True}$ 代表参考答案中关键词的个数。

另外，ATWV 指标是 NIST 评测中一种关键词唤醒任务通用评价指标。它是在每个关键词上平均 TWV 值，可以反映关键词唤醒系统对于不同频次关键词唤醒效果的平均性能，并且与关键词词表的相关性更低。ATWV 的计算公式为：

$$\text{TWV}(\theta) = 1 - [P_{Miss}(\theta) + \beta P_{FA}(\theta)]$$

其中，θ代表关键词检测结果的置信度，在对应置信度θ下：

$$P_{Miss}(\theta) = [\sum_{kw=1}^{K} N_{Miss}(kw, \theta)/N_{True}(kw)]/K$$

$$P_{FA}(\theta) = [\sum_{kw=1}^{K} N_{FA}(kw, \theta)/N_{NT}(kw)]/K$$

$$N_{NT}(kw) = T - N_{True}(kw)$$

其中，$K$ 是关键词总数，$\beta = 999.9$，$T$ 是测试集语音的时长。F1 和 ATWV 指标数值越高，表示系统性能越好。

## 5 实验结果

### 5.1 受限赛道结果

在受限赛道下，表 2 列出了提交系统的 Top10 结果。这些提交系统主要有三类关键词唤醒方法，分别是基于 Conformer 模型[7,8]的端到端关键词唤醒（系统 1/4）基于 Chain 模型[9,10]的 Hybrid 关键词唤醒（系统 5/6/7/8/9/10）和基于端到端和 Hybrid 混合的关键词唤醒方法（系统 2/3）。从结果上看，基于 Conformer 模型的端到端关键词唤醒在三个场景中的效果都要普遍优于基于 Chain 模型的 Hybrid 关键词唤醒方法，而且两种系统融合还能够进一步提升唤醒效果。具体而言，每一类方法的主要创新点如下所示：

在第一类方法中，系统 1 首先通过引入更大的 conformer 主干网络、关键词偏移、音节建模去提升关键词的召回能力，然后使用多级匹配、模糊匹配、声学模型打分的方法确定关键词的唤醒结果；系统 4 则是采用通过 Conformer 模型识别语音内容，然后通过 AC 自动机算法匹配关键词。

在第二类方法中，系统 10 是公开的基线，该基线采用的是基于 CNN-TDNNF 的 Hybrid 模型，并且通过 WFST 解码得到最终的关键词唤醒结果和置信度得分。为了进一步提升 Hybrid 模型效果，很多参数队伍用到了数据增强（系统 5/8）、发音词典和建模音素的选择（系统 5/9）、声学模型结构改进（系统 5/6/9），解码和语言模型改进（系统 5/6/7/9）和输出结果融合（系统 5/6）等创新工作。

在第三类方法中，系统 2 首先采用 ArpaLM 和 RNNLM 同时打分并通过 WFST 解码得到 Hybrid 关键词唤醒结果，然后通过 Conformer 模型得到端到端关键词唤醒结果，最后将两个结果进行合并作为最终的关键

词搜索结果。而系统 3 则是通过基于 CNN-TDNN-LSTM 的 Hybrid 模型和基于不带调拼音的 Conformer 模型来进行最终的输出结果融合。

表 2 受限赛道下测试集结果

| 系统 ID | 框架 | 声学模型 | 长视频 F1 | 长视频 ATWV | 直播 F1 | 直播 ATWV | 短视频 F1 | 短视频 ATWV |
| --- | --- | --- | --- | --- | --- | --- | --- | --- |
| 1 | 端到端 | Conformer | 0.6367 | **0.8488** | **0.8720** | **0.8687** | **0.9103** | 0.9286 |
| 2 | 端到端+Hybrid | Conformer+TDNNF | 0.5889 | 0.7821 | 0.8285 | 0.8453 | 0.8796 | **0.9296** |
| 3 | 端到端+Hybrid | Conformer+CNN-TDNN-LSTM | 0.5986 | 0.7969 | 0.7865 | 0.8349 | 0.8440 | 0.8670 |
| 4 | 端到端 | Conformer | **0.6682** | 0.6883 | 0.8454 | 0.7133 | 0.8863 | 0.7515 |
| 5 | Hybrid | CNN-TDNNF | 0.6132 | 0.7244 | 0.8090 | 0.7816 | 0.8555 | 0.8404 |
| 6 | Hybrid | CNN-TDNNF | 0.4944 | 0.6806 | 0.6614 | 0.7471 | 0.7331 | 0.8121 |
| 7 | Hybrid | TDNNF | 0.5621 | 0.6196 | 0.6101 | 0.5967 | 0.7244 | 0.6885 |
| 8 | Hybrid | CNN-TDNNF | 0.5011 | 0.5189 | 0.6569 | 0.5515 | 0.7274 | 0.6494 |
| 9 | Hybrid | TDNN | 0.3532 | 0.5461 | 0.6214 | 0.6472 | 0.7110 | 0.7292 |
| 10 | Hybrid | CNN-TDNNF | 0.5078 | 0.5482 | 0.5700 | 0.5939 | 0.6682 | 0.6710 |

## 5.2 非受限赛道

在非受限赛道下，表 3 列出了提交系统的 Top4 结果。这些提交系统虽然依然采用的是之前描述的三类主要方法，但是也存在着不少的创新工作。系统 1 在 CN-CELEB 数据集上进行多轮次的半监督学习来提升关键词唤醒的效果；系统 2 和系统 3 则是整理了网上开源的中文数据集来丰富训练数据；系统 4 是直接使用基于外部标注语料的声学模型。

表 3 非受限赛道下测试集结果

| 系统 ID | 框架 | 声学模型 | 长视频 F1 | 长视频 ATWV | 直播 F1 | 直播 ATWV | 短视频 F1 | 短视频 ATWV |
| --- | --- | --- | --- | --- | --- | --- | --- | --- |
| 1 | 端到端 | Conformer | 0.6247 | **0.8495** | **0.8705** | **0.8727** | **0.9080** | 0.9255 |
| 2 | 端到端 | Conformer | 0.6863 | 0.7252 | 0.8660 | 0.7623 | 0.8979 | 0.7941 |
| 3 | 端到端+Hybrid | Conformer+CNN-TDNN-LSTM | 0.5829 | 0.7990 | 0.8107 | 0.8619 | 0.8734 | 0.9372 |
| 4 | 端到端 | Conformer | 0.6761 | 0.7116 | 0.8645 | 0.7420 | 0.9020 | 0.80901 |

## 3 结论

在 VKW 竞赛中，参赛队伍仅用 1500 小时普通话朗读数据和 5 小时目标场景数据，就能够搭建一套高效的关键词唤醒系统。相比官方提供的基线，大部分参赛队伍提交的关键词唤醒系统的效果都有一个明显的提升，有的系统甚至取得了 SOTA 结果。总体而言，端到端系统的关键词唤醒效果已经超越了传统 Hybrid 系统。很多数据增强、建模单元改进、声学模型改进、解码和语言模型改进、结果融合等等方法对关键词唤醒系统依然有进一步提升。在之后的工作中，我们会去探索更高效的关键词唤醒方法，加快关键词唤醒的推理速度，提升关键词唤醒在目标场景的鲁棒性。

## 作者简介

姓　名　袁有根，腾讯科技（北京）有限公司，算法工程师，主要研究语音关键词。E-mail：yougenyuan@tencent.com

姓　名　吕志强，腾讯科技（北京）有限公司，算法工程师，主要研究语音关键词。E-mail：zhiqianglv@tencent.com

姓　名　黄　申，腾讯科技（北京）有限公司，算法工程师，主要研究语音关键词。E-mail：springhuang@tencent.com

姓　名　胡鹏飞，腾讯科技（北京）有限公司，算法工程师，主要研究语音关键词。E-mail：alanpfhu@tencent.com

## 创新点说明

组织了汉语长短视频直播语音关键词竞赛(Video Keyword Wakeup Challenge, VKW)

提供了VKW公开测试集和评测工具

描述了VKW竞赛的基本情况和部分参赛队伍的实验结果